\documentclass[lettersize,journal]{IEEEtran}
\usepackage{amsmath,amsfonts}
\usepackage{algorithmic}
\usepackage{algorithm}
\usepackage{array}
\usepackage[caption=false,font=normalsize,labelfont=sf,textfont=sf]{subfig}
\usepackage{textcomp}
\usepackage{stfloats}
\usepackage{url}
\usepackage{verbatim}
\usepackage{graphicx}
\usepackage{multirow}
\usepackage[noadjust]{cite}
\usepackage{xcolor}
\begin{document}

\title{Advancing Education through Tutoring Systems: A Systematic Literature Review}

\author{Vincent Liu, Ehsan Latif, and Xiaoming Zhai*
\thanks{The authors are with the AI4STEM Education Center, University of Georgia, Athens, GA, 30602, USA. \\
Corresponding Author Email: {\tt\small xiaoming.zhai@uga.edu}  
}}
        % <-this % stops a space



\maketitle

\begin{abstract}
This study systematically reviews the transformative role of Tutoring Systems, encompassing Intelligent Tutoring Systems (ITS) and Robot Tutoring Systems (RTS), in addressing global educational challenges through advanced technologies. As many students struggle with proficiency in core academic areas, Tutoring Systems emerge as promising solutions to bridge learning gaps by delivering personalized and adaptive instruction. ITS leverage artificial intelligence (AI) models, such as Bayesian Knowledge Tracing and Large Language Models, to provide precise cognitive support, while RTS enhance social and emotional engagement through human-like interactions. This systematic review, adhering to the PRISMA framework, analyzed 86 representative studies. We evaluated the pedagogical and technological advancements, engagement strategies, and ethical considerations surrounding these systems. Based on these parameters, Latent Class Analysis was conducted and identified three distinct categories: computer-based ITS, robot-based RTS, and multimodal systems integrating various interaction modes. The findings reveal significant advancements in AI techniques that enhance adaptability, engagement, and learning outcomes. However, challenges such as ethical concerns, scalability issues, and gaps in cognitive adaptability persist. The study highlights the complementary strengths of ITS and RTS, proposing integrated hybrid solutions to maximize educational benefits. Future research should focus on bridging gaps in scalability, addressing ethical considerations comprehensively, and advancing AI models to support diverse educational needs.
% As artificial intelligence continues to reshape education, tutoring systems have emerged as transformative tools for delivering personalized and adaptive learning experiences. This systematic review examines the evolution of tutoring systems, focusing on software-based Intelligent Tutoring Systems (ITS) and Robot Tutoring Systems (RTS). We found that ITS demonstrates exceptional cognitive adaptability, leveraging advanced AI techniques such as Bayesian Knowledge Tracing and large language models to tailor content dynamically. RTS, on the other hand, excels in fostering engagement and motivation through physical interactions and social presence but often relies on external devices for cognitive adaptability. Both systems face challenges in scalability, ethical data management, and accommodating diverse learning styles. Our review also reveals that while ITS has been extensively studied and widely adopted, RTS remains underexplored. Our analysis underscores the potential of integrating ITS and RTS to combine their respective strengths, creating hybrid systems that address both cognitive and emotional needs. Such integration could lead to a more comprehensive educational experience, dynamically adapting to both academic performance and social cues.  Future research should prioritize bridging these gaps to unlock the full potential of tutoring systems in diverse educational settings.
\end{abstract}


\section{Introduction}
\IEEEPARstart{I}{mproving} academic performance remains a global challenge, with many students struggling to achieve proficiency in core academic areas. For instance, a recent study highlighted that 96\% of students in an urban Pennsylvania school performed below proficiency levels, while none reached proficiency at another site \cite{thomas2024improving}. These findings underscore the critical need for innovative educational interventions that address systemic learning gaps.

\textbf{Tutoring Systems}, encompassing both \textit{Intelligent Tutoring Systems (ITS)} and \textit{Robot Tutoring Systems (RTS)}, are emerging as transformative educational technologies. ITS are software-based systems leveraging artificial intelligence (AI) to deliver adaptive, personalized instruction in real time \cite{mousavinasab2021intelligent, kulik2016effectiveness}. RTS extend these capabilities by integrating robotics to provide physical and social interactivity, fostering engagement through tangible interactions \cite{chu2022artificial, zhang2024bring}. Collectively, Tutoring Systems have demonstrated significant potential to enhance learning outcomes by adapting to individual student needs, promoting motivation, and addressing the limitations of traditional instructional methods \cite{ma2014intelligent, feng2021systematic}.

The significance of this study lies in its systematic evaluation of Tutoring Systems, particularly in the context of their impact on student engagement, academic achievement, and pedagogical practices. Research reveals that ITS can achieve learning gains comparable to one-on-one human tutoring \cite{holstein2017intelligent, maity2024generative}, while RTS have shown promise in improving student motivation and social-emotional learning outcomes \cite{ramachandran2019toward, chin2014impact}. The integration of generative AI into ITS further enhances their adaptability, enabling real-time customization of instructional content based on student behavior and feedback \cite{banjade2024empowering, lin2023artificial}. Similarly, RTS provide unique opportunities for collaborative learning and problem-solving, bridging the digital divide and fostering inclusivity in education \cite{yang2019artificial, wang2023effectiveness}.

The evolution of tutoring systems, from rudimentary Computer Assisted Instruction (CAI) programs in the 1970s to today’s advanced ITS and RTS, has transformed personalized learning. However, while technological breakthroughs have enabled unprecedented levels of adaptive instruction, critical questions remain regarding the practical impact on learning outcomes and the ethical implications of these innovations. Addressing these issues is essential for guiding future research and ensuring that educational technologies contribute positively to student development.

Despite their potential, Tutoring Systems face challenges, including ethical concerns related to AI bias, data privacy, and equitable access \cite{swargiary2024impact, robert2024impact}. This study seeks to address these issues by evaluating advancements in AI-driven tutoring, examining their adaptability to real-time inputs, and exploring best practices to ensure their ethical use \cite{akyuz2020effects, ahmadov2024two}.

Below are the research questions address in this study:
\begin{enumerate}
    \item How have tutoring systems evolved from the early CAI models of the 1970s to today’s AI-driven ITS and RTS, and which technological breakthroughs have been most critical in enabling personalized and adaptive learning?
    \item In what specific ways do modern tutoring systems incorporate real-time adaptive mechanisms, and how does this dynamic responsiveness translate into improved student engagement and academic performance across various educational contexts?
    \item What ethical challenges, such as data privacy, informed consent, and algorithmic bias, have emerged alongside the technological advancements in tutoring systems, and what proven strategies have been developed to mitigate these risks while ensuring equitable and responsible use of these tools?
\end{enumerate}

This paper is structured to systematically explore the role of Tutoring Systems in education. The introduction outlines the motivation and objectives of the study, followed by a detailed framework that characterizes Tutoring Systems. The methodology section describes the systematic review process, employing the PRISMA framework to identify and select relevant studies. The results section addresses the research questions, analyzing the impact of Tutoring Systems on educational outcomes, engagement, and ethical considerations. The discussion highlights key pedagogical and technological gaps, identifies integration challenges, and proposes future research directions. The paper concludes with a synthesis of findings and recommendations for advancing the field of Tutoring Systems.
\section{Intelligent Tutoring System (ITS) vs. Robot Tutoring Systems (RTS)}

This section provides a comparative analysis of Tutoring Systems based on their pedagogical effectiveness, adaptability, personalization, interaction methods, engagement strategies, and technical implementation. The comparison highlights their respective strengths and limitations, offering insights into how these systems address diverse educational needs and contexts.

\begin{table*}[htp!]
\centering
\caption{Comparison of ITS and RTS as Subcomponents of Tutoring Systems}
\begin{tabular}{|l|p{6cm}|p{6cm}|}
\hline
\textbf{Dimension} & \textbf{Intelligent Tutoring Systems (ITS)} & \textbf{Robot Tutoring Systems (RTS)} \\ \hline
\textbf{Pedagogical Focus} 
& Cognitive and instructional personalization through adaptive algorithms and AI models. 
& Social and emotional engagement via physical presence, gestures, and verbal communication. \\ \hline

\textbf{Adaptability} 
& High adaptability with real-time adjustments to content, difficulty, and feedback using advanced AI techniques like Bayesian Knowledge Tracing (BKT) and Large Language Models (LLMs). 
& Adaptability through interactive behaviors such as task adjustments and social responses, but primarily pre-programmed and less data-driven. \\ \hline

\textbf{Personalization} 
& Tailors instructional content using detailed student models to address individual learning preferences and weaknesses. 
& Enhances interaction by recognizing students, using names, and offering social cues such as verbal praise and gestures. \\ \hline

\textbf{Interaction} 
& Text-based or conversational interfaces; cognitive and information-focused interactions. 
& Multimodal interactions including physical gestures, verbal communication, and eye contact, mimicking human tutors. \\ \hline

\textbf{Engagement} 
& Cognitive engagement driven by adaptive content and feedback, maintaining mental stimulation. 
& Emotional and social engagement through human-like behaviors, enhancing motivation and relational connection. \\ \hline

\textbf{Scalability} 
& High scalability due to reliance on existing computer infrastructure and cloud-based deployment. 
& Limited scalability due to the need for physical robotic hardware and associated maintenance. \\ \hline

\textbf{Ethical Considerations} 
& Minimal ethical concerns; primary issues involve data privacy and ensuring fair AI-driven decision-making. 
& Higher ethical concerns related to data privacy, long-term human-robot interaction effects, and transparency in AI usage. \\ \hline

\textbf{Deployment Challenges} 
& Easy deployment on existing digital platforms with minimal hardware requirements. 
& Complex deployment requiring significant investment in robotic hardware and integration into physical environments. \\ \hline
\end{tabular}
\label{tab:ITS_RTS_comparison}
\end{table*}

\subsubsection{Pedagogical Effectiveness}

Tutoring Systems exhibit distinct pedagogical strengths. Software-based systems excel in delivering personalized and adaptive instruction by leveraging advanced AI models such as Bayesian Knowledge Tracing and Large Language Models (LLMs) \cite{fancsali2014context}. These systems provide precise feedback and track progress to ensure cognitive engagement, which significantly enhances learning outcomes. However, they often lack the social and emotional engagement that contributes to holistic education.

Conversely, robot-based systems prioritize social and emotional engagement through physical presence, verbal encouragement, and human-like gestures \cite{kanda2004interactive, ramachandran2019toward}. These features are particularly effective for younger learners or those who benefit from interactive and relational learning. While robot-based systems support motivation and engagement, they face challenges in delivering deeply personalized and cognitively adaptive content at the same level as software-based systems.

\subsubsection{Adaptability}

Adaptability is a defining feature of software-based tutoring systems, achieved through dynamic adjustments to content and difficulty levels based on real-time student performance \cite{ines2024twenty}. These systems utilize AI-driven techniques to provide highly individualized support, ensuring students are consistently challenged at appropriate levels.

Robot-based systems demonstrate adaptability primarily through interactive behaviors, such as providing hints or altering task complexity based on student responses \cite{verhelst2024adaptive}. However, this adaptability is often limited to pre-programmed responses, making it less flexible and nuanced compared to software-based systems. While robots excel in maintaining engagement, their cognitive adaptability remains an area for further development.

\subsubsection{Personalization}

Software-based systems personalize learning by analyzing detailed student profiles, including performance data and learning preferences, to tailor content and feedback effectively \cite{fancsali2014context}. This approach ensures a targeted learning experience that addresses individual strengths and weaknesses.

Robot-based systems personalize interactions through social and emotional engagement, such as recognizing students by name and using verbal and non-verbal cues to foster a sense of connection \cite{kanda2004interactive}. However, their personalization is more focused on enhancing relational aspects rather than deeply customizing instructional content, limiting their cognitive personalization capabilities.

\subsubsection{Interaction}

The interaction styles of Tutoring Systems differ significantly. Software-based systems rely on text or conversational interfaces powered by AI models, offering efficient and context-aware cognitive engagement \cite{lieb2024student}. However, their interactions remain screen-based, lacking the immersive and physical elements of human-like interaction.

Robot-based systems provide rich multimodal interactions, combining verbal communication, gestures, and eye contact to create a more personal and engaging learning experience \cite{verhelst2024adaptive, kanda2004interactive}. These physical and social interactions enhance motivation and emotional connection, making them particularly effective for younger learners.

\subsubsection{Engagement}

Engagement in software-based systems is primarily cognitive, driven by adaptive content and immediate feedback that keeps learners mentally stimulated \cite{swargiary2024impact}. While effective in maintaining focus and promoting academic achievement, these systems may lack the emotional resonance required for sustained motivation.

Robot-based systems excel in emotional and social engagement by mimicking human behaviors, offering encouragement, and fostering a sense of companionship \cite{chin2014impact}. This approach increases students’ willingness to participate and enhances their attitudes toward learning, though it may not provide the same depth of cognitive engagement as software-based systems.

\subsubsection{Technical Implementation}

Software-based systems are relatively easy to deploy and scale due to their reliance on existing infrastructure and minimal hardware requirements. Cloud-based solutions further simplify updates and management, making these systems cost-effective and accessible in diverse educational settings.

Robot-based systems face greater deployment challenges, requiring significant investments in robotic hardware and supporting infrastructure. Scalability is limited by the cost and complexity of maintaining additional units. Hybrid approaches, such as combining robots with virtual avatars, offer potential solutions but may compromise the immersive experience robots are designed to provide \cite{zhang2024bring, wang2023effectiveness}.

This comparative analysis underscores the complementary nature of software-based and robot-based Tutoring Systems. While software-based systems excel in cognitive adaptability and scalability, robot-based systems bring unique strengths in emotional and social engagement. Together, these systems can address diverse educational needs, offering opportunities for hybrid solutions that combine their respective advantages.

\section{Framework for Analyzing Tutoring Systems}\label{sec:framework}

\begin{figure}[htp!]
    \centering
    \includegraphics[width=\linewidth]{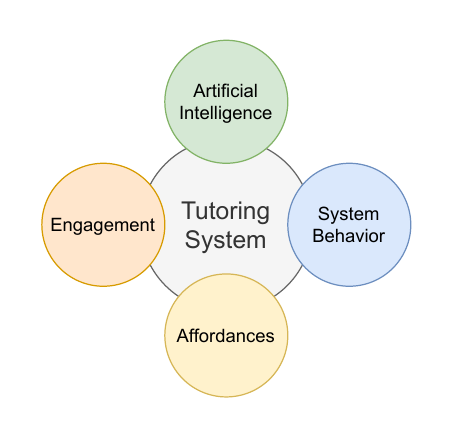}
    \caption{Framework for Tutoring System Analysis}
    \label{fig:framework}
\end{figure}

This review introduces a framework designed to analyze the evolution of tutoring systems across different generations. The framework is structured around four dimensions: Artificial Intelligence, system behavior, engagement, and affordances. To ensure a comprehensive analysis, representative papers from each generation of tutoring systems were carefully selected based on their alignment with these dimensions. These papers, identified through an extensive review of the literature, exemplify the defining characteristics and advancements of their respective generations. By systematically applying this framework to representative studies, the review provides a nuanced evaluation of the technological and pedagogical evolution of tutoring systems, highlighting their strengths, limitations, and potential for future integration (Fig.~\ref{fig:framework}).

To systematically assess the progression of tutoring systems, we developed a framework that categorizes their evolution across different generations. This framework was constructed based on key technological advancements, pedagogical methodologies, and system capabilities observed in the literature. Therefore, we have structured the framework aroud four dimensions: Artificial Intelligence, system behavior, engagement, and affordances. As educational technologies have advanced, new capabilities, such as adaptive learning, AI-driven feedback, and robotic interaction, have emerged, fundamentally altering how students engage with tutoring systems. These shifts have been driven by advancements in computing power, machine learning, and human-computer interaction, yet there is limited work that systematically maps these developments into a structured framework. By introducing this generational model, we provide a historical context that allows for a clearer understanding of how tutoring systems have evolved to address different educational needs. Furthermore, analyzing tutoring systems through this generational lens strengthens the investigation into the impact of AI-driven advancements, their adaptability, and their ethical implications. Situating our findings within this broader trajectory enables a more comprehensive assessment of tutoring systems, ultimately offering insights that can inform future research and development in the field.

\subsection{Artificial Intelligence}
In the context of tutoring systems, Artificial Intelligence encompasses the algorithms and computational models that enable these technologies to deliver personalized and adaptive learning experiences \cite{lin2023artificial, zhang2024bring}. Early tutoring systems relied on foundational techniques, such as concept trees and student models, to map student progress and adapt instruction based on predefined rules \cite{shute199419, hume1996use, draper1992using}. Subsequent advancements introduced Bayesian Knowledge Tracing, which improved predictive capabilities, allowing systems to adjust instructional content dynamically based on real-time student performance \cite{ines2024twenty}.

Modern tutoring systems integrate advanced AI models, such as ChatGPT and other large language models, to create interactive, conversational learning experiences \cite{lieb2024student, verhelst2024adaptive}. These models facilitate real-time feedback and adaptive interactions, tailoring the learning process to individual student needs. This evolution has transformed tutoring systems into responsive, AI-driven tools that emulate many qualities of human tutors, offering highly adaptive and personalized educational experiences \cite{banjade2024empowering}.

\subsection{System Behavior}

System behavior refers to the adaptability of tutoring systems—their ability to modify instructional strategies and content in response to student inputs. Early systems offered limited adaptability, primarily delivering static content with minimal interaction. Over time, these systems evolved to include features such as adjusting problem difficulty, providing personalized feedback, and introducing tailored challenges based on student progress \cite{kanda2004interactive}.

Advanced tutoring systems now leverage real-time analytics to dynamically adapt task complexity, provide hints, and deliver context-sensitive feedback \cite{ramachandran2019toward, verhelst2024adaptive}. These capabilities enable systems to align more closely with students’ learning paces and styles, fostering a personalized educational experience. By mirroring the flexibility of human tutors, modern systems enhance the learning process, supporting improved engagement and outcomes \cite{zhang2024bring}.

\subsection{Engagement}

Engagement in tutoring systems encompasses the strategies and features designed to maintain students’ active involvement in the learning process \cite{swargiary2024impact}. Early systems relied on basic text-based feedback and limited interaction, offering minimal engagement \cite{hume1996use, draper1992using}. As these systems evolved, they incorporated more interactive elements, such as animated feedback \cite{corbett2000instructional}, subgoal scaffolding \cite{corbett2000instructional}, and real-time monitoring of concentration levels \cite{lee2015concentration}.

Contemporary tutoring systems utilize sophisticated engagement methods, including verbal communication, gestures, and physical movements that simulate human-like interactions \cite{verhelst2024adaptive, kanda2004interactive}. These systems can hold natural conversations, use facial expressions, and adapt dynamically to student responses, creating immersive and motivating learning environments. By integrating these advanced features, modern tutoring systems sustain student interest and motivation, making learning experiences more engaging and effective \cite{ramachandran2019toward}.

\subsection{Affordances}

Affordances in tutoring systems refer to the physical and interactive features that enhance student engagement and interaction. Early tutoring systems were primarily software-based, providing limited physical interaction. Over time, systems began incorporating tangible elements, such as virtual avatars or robotic components, to create more immersive learning experiences \cite{yang2019artificial}.

Modern systems leverage physical features like robotic gestures, digital representations, and virtual environments to foster engagement and accessibility. For example, robotic tutors can use movement and physical presence to enhance interaction, particularly for younger learners or those who benefit from kinesthetic learning styles \cite{wang2023effectiveness, chin2014impact}. These affordances bridge the gap between digital and physical learning, offering diverse and effective educational tools that cater to a wide range of learning needs \cite{zhang2024bring}.

\section{Method}\label{sec:method}

\begin{figure}[H]
    \centering
    \includegraphics[width=\linewidth]{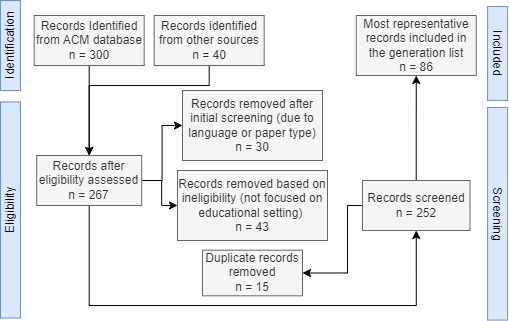}
    \caption{Tutoring Systems PRISMA review flowchart}
    \label{fig:PRISMA_Image}
\end{figure}

This systematic review adheres to the Preferred Reporting Items for Systematic Reviews and Meta-Analyses (PRISMA) guidelines, proposed by Moher et al. \cite{moher2015preferred}, to ensure a transparent, replicable, and comprehensive approach to data collection and analysis. The PRISMA framework facilitated a structured process for identifying, screening, and selecting studies related to Tutoring Systems. The method comprises the following stages: identification, screening, eligibility, and inclusion, as depicted in Figure~\ref{fig:PRISMA_Image}.

\subsection{Search Strategy}

To ensure transparency and replicability, this study adopted the PRISMA framework. As shown in the flowchart in Figure~\ref{fig:PRISMA_Image}, we followed a systematic review process including overview of identification, screening, eligibility, and inclusion stages. To ensure a robust and comprehensive search, multiple bibliographic databases were used, including ACM, JSTOR, ResearchGate, BERA, ScienceDirect, Taylor \& Francis, arXiv, ERIC, IOSPress, MDPI, EPFL, ISEDJ, and the American Psychological Association. Search queries focused on titles, abstracts, and keywords, using combinations of terms such as "Tutoring Systems," "adaptive learning," "AI in education," and "personalized learning."

Table~\ref{table:search_criteria} summarizes the search criteria applied to filter the studies. The initial search yielded a total of 340 records, ensuring inclusivity across article types, including research articles, journals, conference papers, and short papers.

To capture the full evolution of tutoring systems, this review spans a broad historical range from the early 1970s to the present. The earliest paper identified in our search, “CAI supports individualized learning programs in Montgomery County (Maryland) Public Schools” (1972), represents one of the pioneering works in computer-assisted instruction. By including studies from 1972 through 2024, we ensure that our analysis covers the inception of individualized learning technologies, the emergence and growth of ITS and RTS, and the latest innovations incorporating LLMs. This comprehensive time frame allows us to trace the technological, pedagogical, and ethical transformations that have shaped tutoring systems over more than five decades.

\begin{table}[H]
\centering
\caption{Search Criteria}
\label{table:search_criteria}
\begin{tabular}{|l|p{5cm}|}
\hline
\textbf{Components} & \textbf{Examples}\\
\hline
Bibliography Databases & ACM, JSTOR, ResearchGate, BERA, ScienceDirect, Taylor \& Francis, arXiv, ERIC, IOSPress, MDPI, EPFL, ISEDJ, American Psychological Association \\
\hline
Article Types & Research Articles, Short Papers, Journals, Conference Papers \\
\hline
Searched on & Title, Keywords, Abstract \\
\hline
Sorting on Returns & Relevance and Time Period \\
\hline
Language & English \\
\hline
\end{tabular}
\end{table}

\subsection{Inclusion and Exclusion Criteria}

A systematic approach was used to establish inclusion and exclusion criteria, as summarized in Table~\ref{table:criteria}. Studies were included if they were in English, focused on educational settings, and involved experimentation or review of Tutoring Systems. Non-English documents, papers without educational relevance, or those without sufficient experimentation were excluded. Additional criteria ensured the inclusion of studies explicitly addressing Tutoring Systems with substantial educational relevance.

\begin{table}[htp!]
\centering
\caption{Inclusion and Exclusion Criteria for Tutoring Systems}
\label{table:criteria}
\begin{tabular}{|p{4cm}|p{4cm}|}
\hline
\textbf{Inclusion} & \textbf{Exclusion} \\
\hline
Documents in English & Documents not in English \\
\hline
Focuses on educational settings & Not focused on educational settings \\
\hline
Papers with experimentation or reviews & Papers without experimentation \\
\hline
Studies addressing Tutoring Systems with significant teaching/learning focus & Studies with loose or no mention of Tutoring Systems relevance \\
\hline
Doc types: Research articles, journals, conference papers, short papers & Other doc types: Book reviews, posters, etc. \\
\hline
\end{tabular}
\end{table}

\subsection{Study Selection Process}

\textbf{Identification:} An initial search yielded 340 records, with 290 retrieved from the ACM Digital Library and 50 identified through additional searches on Google Scholar and Bing, following these procedures:

\textbf{Screening:} After duplicate removal, 252 unique records remained. This phase eliminated 15 duplicate records. Titles and abstracts were screened for relevance to Tutoring Systems in educational settings.

\textbf{Eligibility:} The eligibility evaluation removed 30 papers due to language barriers or document type incompatibilities (e.g., book reviews or editorials). An additional 43 records were excluded for lacking direct relevance to Tutoring Systems in educational contexts. Consequently, 179 records proceeded to the next stage.

\textbf{Inclusion:} After full-text assessment, 86 papers were included in the final review. These studies were deemed the most representative of their respective domains, aligned with the research objectives, and suited to the analytical framework.

While the review focuses on English-language publications, this criterion may have excluded relevant studies published in other languages. Additionally, the reliance on digital databases may introduce bias against non-digitized or regionally focused studies. This method ensures a thorough exploration of Tutoring Systems, providing a solid foundation for analyzing their impact on education.

\subsection{Data Analysis} 
A systematic data extraction template was developed to according to our analytical framework. Key data points including but not limited to the study objectives, methodologies, sample sizes, educational settings, technologies employed, evaluation metrics, and findings, were coded by the authors. Interater reliability were monitored to make sure that the data coded were reliable. The extracted data were synthesized to identify trends, compare approaches, and highlight gaps in the current literature.

The methodological rigor ensures that this review captures a diverse range of perspectives on Tutoring Systems. By leveraging multiple databases and applying stringent inclusion criteria, the review provides a comprehensive overview of the field, highlighting both advancements and areas requiring further exploration.

\subsection{Variables Measured}

Ethics in tutoring systems is categorized into three levels: no ethical concerns, minimal ethical concerns, and high ethical concerns. Ethical concerns can range from data privacy to fears of abusing the AI to cheat. Papers classified as having no ethical concerns either explicitly state the absence of ethical issues or do not identify any potential risks associated with their tutoring system. For example, \cite{verhelst2024adaptive} does not mention any ethical concerns, and the system described does not present any apparent ethical risks. Papers with minimal ethical concerns acknowledge some potential issues but provide solutions or safeguards to address them. For instance, \cite{dong2023build} highlights a feature allowing users to delete chat history, offering a degree of control over data privacy. In contrast, papers with significant ethical concerns identify more serious risks, such as overreliance on AI or its potential for generating misleading information. A clear example is \cite{kumar2023impact}, which warns readers about the dangers of AI hallucinations potentially misleading students.

Adaptability refers to a system's ability to adjust instructional strategies in real time, linking closely to the dimensions of system behavior and AI. Papers are categorized as either having adaptive or fixed difficulty. Systems with adaptive difficulty, such as \cite{king2021intelligent}, dynamically modify teaching strategies and content based on student performance, creating a personalized learning experience. In contrast, systems with fixed difficulty, like \cite{kanero2021even}, function more as one-way communication tools. These systems present linear content and difficulty levels without incorporating any input from the student, limiting their ability to respond to individual needs.                                             

Interaction tools can significantly shape engagement and affordances in tutoring systems, spanning modalities such as computer software, tablet devices, and physical robots. Systems using computers as their primary interaction medium typically rely on software that acts as the student's tutor. For instance, in \cite{fancsali2014context}, students used software to learn mathematics, with the tutor dynamically personalizing content to meet individual student needs in real time. Similarly, systems employing tablet-like devices often integrate robots as complementary tools to enhance communication. For example, \cite{verhelst2024adaptive} describes a system where a tablet presents questions to students while a robot provides positive and encouraging feedback based on their answers. In contrast, systems that prioritize physical robots as the primary interaction tool offer more hands-on engagement. These robots often leverage gestures, movements, voices, and lifelike physical features to create immersive experiences. An example can be found in \cite{kanda2004interactive}, where robots used a diverse range of gestures and movements to interact with students physically.
The appearance of a tutoring system can also play a role in shaping the overall student experience. These systems can be categorized into non-anthropomorphic, semi-anthropomorphic, and fully anthropomorphic designs. Non-anthropomorphic systems are typically computers or tablet-like devices, which lack human-like features. Semi-anthropomorphic systems incorporate some human-like aspects but are missing key components. For example, \cite{verhelst2024adaptive} describes a robot with a human-like head and shoulders but without a full body. In contrast, fully anthropomorphic robots closely resemble human figures. An example is \cite{lekova2022making}, which uses a NAO robot that resembles a human more closely, with features designed to be friendly and approachable.

Feedback mechanisms, such as text-based or verbal responses, play a crucial role in driving engagement and positively influencing a student’s attitude toward the system. Feedback can be categorized into two main types: text-based and verbal. Text-based systems deliver feedback through written messages, typically displayed on a computer or tablet screen. For example, \cite{jiang2024beyond} describes a system that provides personalized feedback, including knowledge explanations, error analysis, and review suggestions, all conveyed through text on the student's screen. In contrast, verbal feedback systems use spoken responses, often delivered by a robot, to communicate with the student. These systems often incorporate encouraging language to motivate students and offer explanations through a different medium. An example of this is \cite{verhelst2024adaptive}, where the robot repeats the words displayed on the tablet, providing students with an additional layer of interaction and learning beyond just text on the screen.

AI techniques like Bayesian Knowledge Tracing (BKT) and Large Language Models (LLMs) are at the heart of these tutoring systems. BKT is a probabilistic model used in educational technology to track and predict a student's knowledge over time and can be seen in \cite{pardos2023oatutor} where they use BKT to predict the student’s knowledge level and adjust its content accordingly. LLMs are advanced AI systems trained on vast text data to understand and generate human language. In intelligent tutoring systems, LLMs provide personalized feedback, facilitate natural language interactions, generate adaptive content, and help students understand concepts through detailed explanations and dynamic dialogues. This can be seen in \cite{alsafari2024towards} where LLMs provide dynamic, personalized, and content-aware responses.

\section{Findings}\label{sec:findings}

% RQ 1: How many categories of tutoring systems in the literature?
% RQ 2: What are the key feature of each category of the tutoring systems?

\subsection{Latent Class Analysis of Features of Tutoring Systems}
Using categorical variables such as scalability, ethics, adaptability, device used, appearance, feedback, and AI techniques, we conducted Latent Class Analysis (LCA). Although the Bayesian Information Criterion (BIC) and Akaike Information Criterion (AIC) suggest that models with less latent classes provide better statistical fit, we believe that the 3-class model is the most appropriate for our scenario. This model strikes a balance between interpretability and capturing distinct characteristics of the latent classes. Adding more or reducing classes introduces complexity without providing substantial additional insight for this specific context. 

The three primary classes are: Tutoring Systems relying on computer-based environments (ITS), Verbal Tutoring Systems using robot-based interactions, and Multimodal Tutoring Systems integrating multiple interaction modes. Each class exhibits distinct characteristics that align with specific educational affordances and pedagogical applications. This classification provides insights into the design choices and capabilities of tutoring systems and their impact on learning outcomes, engagement, and adaptability.By aligning these findings with the proposed framework, the analysis highlights how the dimensions of AI, system behavior, engagement, and affordances are shaping modern tutoring systems.

\subsection{Categorical Features of Tutoring Systems}
 This section articulated the  key features of each latent class.

\subsubsection{Class 1: Computer-Based Tutoring Systems (ITS)}  
Computer-based tutoring systems are characterized by non-anthropomorphic appearances and primarily text-based feedback. They rely heavily on AI techniques such as BKT and LLMs to deliver adaptive and personalized learning experiences \cite{shute199419, verhelst2024adaptive}. These systems demonstrate high adaptability, with a strong emphasis on real-time adjustments to content difficulty and feedback based on student performance. However, their scalability is moderate, and ethical concerns related to data privacy are generally less emphasized.
An example of an ITS that strongly fits this category is \cite{eryilmaz2020development}, where the system was entirely text-based, with students interacting solely through their computers. The system employed Bayesian Knowledge Tracing (BKT) to track student progress and dynamically adjust content to meet individual needs and difficulty levels. The computer interface was non-anthropomorphic, and the paper raised no ethical concerns.

\subsubsection{Class 2: Verbal Tutoring Systems}  
Verbal tutoring systems are typically robot-based, with semi-anthropomorphic appearances and a focus on visual and verbal feedback. These systems enhance engagement by leveraging physical interactions and social behaviors, such as gestures and verbal cues, to create a more interactive learning environment \cite{ramachandran2019toward, kanda2004interactive}. Although verbal systems demonstrate some adaptability, their reliance on pre-programmed content and external devices, such as tablets, limits their scalability and cognitive adaptability compared to computer-based systems. Ethical concerns are more prominent in this class, given the reliance on sensory data and potential risks associated with prolonged human-robot interactions.
An example of a system that exemplifies the verbal tutoring system category is \cite{helal2024robotic}, where the robot speaks and displays questions for the student to answer. The robot has limited mobility, with slight movements in its head, arm, and torso, making it semi-anthropomorphic. The system adapts dynamically by adjusting its questions in real time, using Large Language Models (LLMs) as its underlying AI technique. The paper also raises significant ethical concerns regarding LLM hallucinations, cautioning students about relying on them for factual information.

\subsubsection{Class 3: Multimodal Tutoring Systems}  
Multimodal tutoring systems integrate advanced AI with highly scalable and interactive features. They employ anthropomorphic designs and verbal feedback, fostering a highly engaging and immersive learning experience \cite{zhang2024bring, wang2023effectiveness}. These systems utilize a variety of AI techniques, including LLMs and other advanced models, to adapt dynamically to student inputs. Multimodal systems excel in engagement and adaptability, offering a combination of verbal, visual, and physical interactions. However, their high complexity raises ethical concerns related to data privacy, transparency, and equitable access.
An example of a system that highlights a multimodal tutoring system is \cite{schodde2017adaptive} where the robot is fully anthropomorphic and equipped with a wide range of gestures and actions that make the NAO feel lifelike. The system uses BKT and can adapt its question in real time similar to other systems. The system did not have any ethical concerns and the feedback was a mix of verbal commands, gestures, and text from and external device.

\begin{table*}[ht!]
\centering
\caption{Latent Class Analysis: Characteristics of Tutoring System Classes}
\begin{tabular}{|l|l|l|c|c|c|}  
\hline
\textbf{Dimension} & \textbf{Variable} & \textbf{Category} & \textbf{Computer-Based} & \textbf{Verbal} & \textbf{Multimodal} \\ 
\hline
\multirow{3}{*}{System Behavior} 
    & \textbf{Adaptability}   
        & Adaptive difficulty   & 40\% & 20\% & 40\% \\ 
    &                       & Fixed difficulty      & 50\% & 50\% & 30\% \\ 
    &                       & Other                 & 10\% & 30\% & 30\% \\ 
    \hline
\multirow{6}{*}{Affordances} 
    & \textbf{Interaction Tool} 
        & Computer              & 100\% & 0\%  & 10\% \\ 
    &                       & Tablet                & 0\%   & 0\%  & 60\% \\ 
    &                       & Physical Robot        & 0\%   & 100\% & 20\% \\ 
    & \textbf{Appearance}    
        & Non-anthropomorphic   & 100\% & 0\%   & 0\%  \\ 
    &                       & Semi-anthropomorphic  & 0\%   & 40\%  & 20\% \\ 
    &                       & Full anthropomorphic  & 0\%   & 40\%  & 80\% \\ 
    \hline
\multirow{2}{*}{Engagement} 
    & \textbf{Feedback}      
        & Text-based            & 90\%  & 50\%  & 0\%  \\ 
    &                       & Verbal                & 10\%  & 50\%  & 100\% \\ 
    \hline
\multirow{6}{*}{Artificial Intelligence} 
    & \textbf{AI Technique}   
        & BKT                  & 20\% & 0\%   & 40\% \\ 
    &                       & LLM                  & 70\% & 60\%  & 40\% \\ 
    &                       & Other                & 10\% & 40\%  & 20\% \\
    & \textbf{Responsible AI}         
        & Ethical concerns      & 20\% & 40\% & 40\% \\ 
    &                       & Minimal concerns      & 20\% & 20\% & 0\%  \\ 
    &                       & No concerns           & 60\% & 40\% & 60\% \\
\hline
\end{tabular}
\label{table:latent_classes}
\end{table*}

\begin{table}[ht!]
\centering
\caption{Comparison of AIC and BIC for Different Numbers of Latent Classes}
\begin{tabular}{|c|c|c|}
\hline
\textbf{Number of Classes (k)} & \textbf{AIC}      & \textbf{BIC}      \\ \hline
1                              & -134.54           & -128.57           \\ \hline
2                              & -786.13           & -773.19           \\ \hline
3                              & -1075.37          & -1055.45          \\ \hline
4                              & -1168.59          & -1141.71          \\ \hline
5                              & -1255.41          & -1221.55          \\ \hline
\end{tabular}
\label{table:aic_bic_comparison}
\end{table}

This classification of tutoring systems demonstrates the diverse ways in which educational technologies can be tailored to meet specific learning needs. The latent class analysis reveals the strengths and limitations of each class, providing insights into how Tutoring Systems can address scalability, ethical considerations, adaptability, and engagement. By aligning these findings with the proposed framework, this analysis highlights areas for future development, particularly in enhancing cognitive adaptability and addressing ethical challenges.

\subsubsection{Advancements in Artificial Intelligence}

Tutoring Systems have witnessed substantial advancements in Artificial Intelligence, enabling more personalized and adaptive learning experiences \cite{lin2023artificial, zhang2024bring}. Early systems relied on fundamental AI techniques such as concept trees and Bayesian Knowledge Tracing to track student progress and adjust content delivery \cite{shute199419}. Recent systems have adopted sophisticated models, including large language models (LLMs), to enhance responsiveness and interactivity \cite{verhelst2024adaptive}. These AI-driven models allow systems to dynamically tailor instructional content, deliver real-time feedback, and facilitate natural language interactions.

Modern Tutoring Systems also integrate Natural Language Processing (NLP) technologies, which support conversational tutoring and improve the overall student experience \cite{ramachandran2019toward}. For example, LLM-based "tutor models" provide tailored instruction, while "feedback models" offer context-sensitive responses, significantly improving engagement and learning outcomes \cite{maity2024generative}. These advancements position Tutoring Systems as powerful tools for adaptive learning, with the capability to emulate many aspects of human tutoring.

\subsubsection{System Behavior and Adaptability}

The adaptability of Tutoring Systems is a cornerstone of their effectiveness. AI-powered algorithms such as Bayesian Knowledge Tracing and LLMs enable these systems to modify instructional strategies in real-time, responding to individual student needs \cite{swargiary2024impact}. For instance, modern systems dynamically adjust task difficulty, provide immediate feedback, and introduce new challenges based on a student’s progress, ensuring an optimal balance of support and challenge \cite{verhelst2024adaptive}.

The physical and social affordances of Tutoring Systems further enhance their adaptability. Robotic components within some systems enable real-time interactions, such as gestures, verbal praise, and physical task demonstrations, fostering engagement and motivation \cite{ramachandran2019toward, wang2023effectiveness}. However, cognitive adaptability—adjusting instructional content based on complex real-time analysis—remains more advanced in AI-driven software-based systems compared to robotic ones. While robots excel in maintaining engagement, their instructional capabilities often rely on external devices or pre-programmed content \cite{zhang2024bring}.

Studies comparing student performance with and without Tutoring Systems have demonstrated their potential to enhance learning outcomes and engagement. However, while software-based systems show significant improvements in academic achievement, the direct cognitive impact of robotic systems remains an area for further exploration.

\subsubsection{Engagement through Interactive Features}
Engagement is a critical dimension of Tutoring Systems, supported by features that actively involve students in the learning process \cite{swargiary2024impact}. Early systems provided basic interaction through static text and simple playback functions, which evolved into more dynamic tools incorporating animated feedback, verbal communication, and real-time monitoring of student attention \cite{corbett2000instructional, lee2015concentration}.

Modern systems integrate sophisticated methods, such as conversational AI and robot-driven gestures, to create immersive and interactive learning environments \cite{verhelst2024adaptive}. These features maintain student interest and motivation over extended periods, making the systems more effective than traditional static instructional methods \cite{chin2014impact, kanda2004interactive}. Advanced Tutoring Systems now employ subtle expressions, such as gazing or facial cues, which simulate human behavior, further enhancing the quality of engagement and aligning with diverse student needs.

\subsection{Ethical Considerations and Responsible AI Use}

The rapid adoption of AI-powered Tutoring Systems raises critical ethical concerns, particularly regarding data privacy, security, and fairness \cite{robert2024impact}. Despite collecting extensive student data, many studies lack a thorough examination of how this sensitive information is managed and protected. This highlights a significant gap in the literature and underscores the need for robust ethical frameworks.

To address these challenges, key measures for responsible AI use include implementing data encryption protocols, conducting regular security audits, and establishing stringent access controls to safeguard student data \cite{eden2024integrating}. Transparency is also essential, with educational institutions and developers required to clearly communicate data collection and usage practices to students and guardians. Mitigating AI biases through diverse training datasets and fairness-aware algorithms is equally critical. Regular audits and human oversight are necessary to ensure equitable treatment and maintain accountability.

Moreover, governance frameworks with oversight committees can support the ethical deployment of Tutoring Systems, ensuring compliance with established data protection standards and promoting trust in AI-driven education \cite{maity2024generative}. By integrating these measures, Tutoring Systems can foster a secure, equitable, and effective learning environment.

This review highlights the transformative potential of Tutoring Systems, driven by advancements in Artificial Intelligence, enhanced adaptability, and interactive engagement features. While these systems offer promising solutions to educational challenges, ethical considerations remain a crucial area for development. Future research should focus on bridging gaps in robotic cognitive adaptability, addressing ethical concerns comprehensively, and exploring long-term impacts on student learning outcomes.

\section{Discussion}

This study highlights the transformative potential of Tutoring Systems in addressing educational challenges through advanced Artificial Intelligence, adaptability, and interactive engagement. Software-based systems demonstrate exceptional cognitive adaptability and scalability, leveraging AI models such as Bayesian Knowledge Tracing and Large Language Models (LLMs) to deliver personalized, adaptive learning experiences \cite{fancsali2014context, verhelst2024adaptive}. These systems are particularly effective in tracking student progress, tailoring content, and providing real-time feedback, which supports robust cognitive engagement and improved learning outcomes.

Robot-based systems, on the other hand, excel in fostering emotional and social engagement through human-like behaviors such as gestures, verbal encouragement, and physical interactions \cite{ramachandran2019toward, kanda2004interactive}. These systems are highly effective in maintaining student motivation and creating immersive learning environments, especially for younger learners or those requiring more relational approaches \cite{zhang2024bring}. Despite these strengths, both systems exhibit distinct limitations, such as the lack of holistic engagement in software-based systems and the scalability challenges faced by robot-based systems.

The integration of software- and robot-based Tutoring Systems offers significant potential to address these limitations. By combining the cognitive adaptability of software-based systems with the social and emotional engagement of robot-based systems, integrated solutions could provide a more comprehensive educational experience. Such systems could dynamically adapt content and interactions to both cognitive performance and emotional states, promoting deeper learning and sustained motivation \cite{wang2023effectiveness}.

Despite these promising findings, several gaps remain that limit the full potential of Tutoring Systems. Future research should address these gaps.

\textbf{Pedagogical Gaps:}  
While software-based systems excel in cognitive engagement, their reliance on text-based or screen interactions fails to accommodate diverse learning styles, particularly kinesthetic and auditory learners \cite{swargiary2024impact}. Robot-based systems, although effective in fostering engagement, often lack the cognitive adaptability and precision required for delivering highly personalized instructional content \cite{verhelst2024adaptive}. Both systems need to adopt a more integrated pedagogical approach that addresses the cognitive, emotional, and experiential dimensions of learning.

\textbf{Technological Gaps:}  
Data privacy and security remain critical concerns, with many Tutoring Systems collecting extensive student data but lacking robust protocols for its protection \cite{eden2024integrating, swargiary2024impact}. The integration of LLMs, while enhancing adaptability, introduces risks such as potential misuse for academic dishonesty, necessitating improved monitoring mechanisms \cite{maity2024generative}. In robot-based systems, the reliance on dual-system designs (robots as adjuncts to tablets) highlights a lack of fully integrated AI capabilities, limiting their potential as autonomous educational tools \cite{ramachandran2019toward}.

\textbf{Scalability and Accessibility Gaps:}  
Software-based systems, while scalable, face challenges in reaching under-resourced regions due to disparities in technological infrastructure and socio-economic barriers \cite{zhang2024bring}. Robot-based systems encounter even more pronounced scalability challenges due to the high costs and logistical complexities of deploying robotic hardware in diverse educational settings \cite{wang2023effectiveness}. Addressing these gaps requires the development of cost-effective solutions that ensure equitable access to advanced educational tools.

\textbf{Ethical Considerations:}  
Both types of systems need to prioritize ethical considerations, including data transparency, fairness, and accountability. The long-term social implications of human-robot interaction in education remain underexplored, raising concerns about dependency, equity, and the potential impact on student development \cite{ramachandran2019toward, kanda2004interactive}.

Addressing these gaps will require a multifaceted approach:
\textbf{Holistic Integration:} Developing systems that integrate the cognitive strengths of software-based systems with the social and emotional capabilities of robot-based systems can create balanced learning environments tailored to diverse needs.

\textbf{Enhanced AI Models:} Improving the capabilities of AI to operate seamlessly within robot-based systems will enhance their adaptability and instructional capabilities, moving beyond adjunctive designs \cite{verhelst2024adaptive}.

\textbf{Scalable Solutions:} Designing cost-effective, accessible systems for deployment in under-resourced settings can reduce disparities in educational technology access.

\textbf{Ethical Frameworks:} Establishing transparent guidelines for data collection, usage, and security will build trust and ensure equitable treatment of all students.

By addressing these challenges, future Tutoring Systems can become more effective, equitable, and transformative, significantly contributing to the advancement of global education.

\section{Conclusion}
The evolution of tutoring systems highlights their transformative potential in advancing educational outcomes through personalized and adaptive learning experiences. Software-based systems, exemplified by ITS, excel in cognitive adaptability, leveraging advanced AI techniques to deliver tailored instruction. Conversely, RTS brings unique strengths in engagement and motivation through physical presence and human-like interactions, making them particularly effective for relational and immersive learning. This review suggests that integrating ITS and RTS offers a promising path forward. By combining the cognitive precision and scalability of ITS with the interactive and engaging qualities of RTS, hybrid systems could address a broader range of student needs. Such systems have the potential to dynamically adapt to both academic performance and social-emotional cues, providing tailored support and feedback in real time. This integration could enhance learning outcomes, particularly for students who benefit from multi-sensory and experiential learning environments.

\section*{Acknowledgment}
This study secondary analyzed data from projects supported by the Institute of Education Sciences (Grant Number R305C240010, PI Zhai). The authors acknowledge the funding agencies and the project teams for making the data available for analysis. The findings, conclusions, or opinions herein represent the views of the authors and do not necessarily represent the views of personnel affiliated with the funding agencies.

\section*{Declaration of generative AI and AI-assisted technologies in the writing process}
During the preparation of this work the author(s) used ChatGPT in order to check grammar and polish the wordings. After using this tool/service, the authors reviewed and edited the content as needed and take full responsibility for the content of the publication.

% \section*{Data availability statement}
% The authors confirm that the data supporting the findings of this study are available within the article.


\bibliographystyle{IEEEtran}
\bibliography{references}

% \newpage

 

% \end{IEEEbiography}



% \vfill

\end{document}